\documentclass[a4paper,prd,preprintnumbers,twocolumn,superscriptaddress,nofootinbib,amsmath,amssymb]{revtex4-2}
\usepackage[utf8]{inputenc}
\usepackage{hyperref}
\hypersetup{colorlinks=true,linkcolor=blue,citecolor=cyan}

\usepackage{graphicx}
\usepackage{dcolumn}
\usepackage{bm}
\usepackage{color}
\usepackage{enumitem}
\usepackage{amsmath}
\usepackage{amssymb}

\begin{document}
\def\cred#1{\textcolor{red}{#1}}
\def\cblue#1{\textcolor{blue}{#1}}

\newcommand{\mr}[1]{\textcolor{blue}{\bf Mirzabek: #1}}

	\title{Shadows and weak gravitational lensing for black holes\\ within Einstein-Maxwell-scalar theory}
	
	\author{Ahmad Al-Badawi}
	\email{ahmadbadawi@ahu.edu.jo}
	\affiliation{Department of Physics, Al-Hussein Bin Talal University, P. O. Box: 20, Ma’an 71111, Jordan}
	
	\author{Mirzabek Alloqulov}
	\email{malloqulov@gmail.com}
	\affiliation{Institute of Fundamental and Applied Research, National Research University TIIAME, Kori Niyoziy 39, Tashkent 100000, Uzbekistan}
	\affiliation{University of Tashkent for Applied Sciences, Str. Gavhar 1, Tashkent 100149, Uzbekistan}
	
	\author{Sanjar Shaymatov}
	\email{sanjar@astrin.uz}
	\affiliation{Institute for Theoretical Physics and Cosmology, Zhejiang University of Technology, Hangzhou 310023, China}
	\affiliation{Institute of Fundamental and Applied Research, National Research University TIIAME, Kori Niyoziy 39, Tashkent 100000, Uzbekistan}
	\affiliation{University of Tashkent for Applied Sciences, Str. Gavhar 1, Tashkent 100149, Uzbekistan}
	\affiliation{Western Caspian University, Baku AZ1001, Azerbaijan}
	
	\author{Bobomurat Ahmedov}
	\email{ahmedov@astrin.uz}
	\affiliation{Institute of Fundamental and Applied Research, National Research University TIIAME, Kori Niyoziy 39, Tashkent 100000, Uzbekistan}

	%

	\date{\today}
	\begin{abstract}
		In this study, we investigated the optical properties of charged black holes within the Einstein-Maxwell-scalar (EMS) theory. We evaluated the shadow cast by these black holes and obtained analytical solutions for both the radius of the photon sphere and that of the shadow. We observed that black hole parameters $\gamma$ and $\beta$ both influence the shadow of black holes. We also found that the photon sphere and shadow radius increase as a consequence of the presence of the parameter $\gamma$. Interestingly, the shadow radius decreases first and then remains unchanged owing to the impact of the parameter $\beta$. Finally, we analyzed the weak gravitational lensing and total magnification of lensed images around black holes. We found that the charge of the black hole and the parameter $\beta$ both have a significant impact, reducing the deflection angle. Similarly, the same behavior for the total magnification was observed, also as a result of the effect of the charge of the black holes and the parameter $\beta$.
	\end{abstract}
	
	\maketitle
	\footnotesize
	\section{Introduction}
	
	In general relativity (GR), black holes have been considered to date as a generic result of finding exact analytical solutions to the field
	equations. Recent observations associated with gravitational waves \cite{Abbott16a,Abbott16b} and with the first image of the elliptical M87 galaxy as a supermassive black hole observed by the Event Horizon Telescope (EHT) collaboration \cite{Akiyama19L1,Akiyama19L6,Hendi_2023} have proven the existence of black holes in the universe. Therefore, these modern observations in connection with black holes have become increasingly important to reach a deeper understanding of their attractive nature and to examine the spacetime geometry in the frame of various theories of gravity.
	
	The EHT collaboration~\cite{Akiyama19L1,Akiyama19L6} published a black hole image in 2019. Near black holes, light can be strongly deflected, even traveling in a circular path. Because of this strong deflection, and the fact that no light comes out of a black hole, a black hole appears as a dark disc in the sky; this disc is known as the black hole shadow. The shadow of a nonrotating black hole is a circular disc. Synge \cite{synge} conducted the first study of light deflection around a Schwarzschild black hole, and Luminet \cite{luminet} simulated a shadow photograph of the black hole. In this direction, shadow analysis was first considered around a Kaluza-Klein rotating dilaton black hole (see for example \cite{Amarilla:2013sj}). Later, shadow analysis was extended to Einstein-Maxwell-Dilaton-Axion (EMDA) black holes and naked singularities \cite{Wei:2013kza}.  The size and shape of the shadow depend on the mass, charge, and angular momentum. We found that, for fixed values of these parameters, the shadow is slightly larger and less deformed than that of its Kerr-Newman counterpart. Shadow analysis plays an important role in probing black hole spacetime geometry and accretion models in EMDA supergravity theory \cite{Jana_2023,R_der_2023}. Similarly, the null-like geodesics around compact objects can also act as the best tools to classify the spacetime structure of charged static dilaton black holes in Einstein-Maxwell-dilaton gravity \cite{Heydari_Fard_2022}. Furthermore, EMDA gravity can also be regarded as an important tool in addressing the late time acceleration of the universe. Thus, it is worth exploring aspects of such gravity model to explain astrophysical observations, including shadows of M87$^\star$ and Sgr A$^\star$ black holes \cite{Sahoo:2023czj}. The shadows of rotating charged black holes with a scalar dilaton field in the environment surrounding plasma medium were also investigated~\cite{Badia:2022phg}. Moreover, the parametrized axially symmetric black holes can include infinite number of deformation parameters that can alter the black hole geometry, thus giving rise to new methods for shadow calculations \cite{Younsi_2016}. This approach has been employed for analyzing the impact of some of these deformation parameters on the black hole shadow in order to enhance our understanding about these objects~\cite{Konoplya_2021}.
	As a result of the detection of black hole shadows, many researchers have worked on theoretical modeling of black hole shadows in recent years \cite{Singh,Konoplya2019,hunting,kumar2020,Afrin21a,abdujabbarov16,zhang,Konoplya19PRD,Atamurotov21JCAP,Mustafa22CPC,Tsukamoto_2014,Tsukamoto_2018,Eslam_Panah_2020,Olmo_2023,asuküla2023spherically}. It should also be noted that black hole shadows have been extended to alternative compact object models, e.g., scalar boson and Proca stars, by adapting analytical fittings of numerical solutions (see for example \cite{Rosa_2022,Rosa_2023,Rosa_2023_2}).  
	
	In GR, it is well-known that gravitational lensing is described by the deflection angle of the light ray, which deviates from its original path owing to a distant source regarded as a massive compact object. Hence, the impact of the background spacetime on gravitational lensing has been a fascinating research topic in the astrophysical field. Interestingly, the first experiment conducted for GR testing was based on the gravitational lensing effect (see for example \cite{Eddington1919GL}). Gravitational lensing has since been considered one of the fundamental tests to collect information concerning distance sources and compact objects (e.g., black holes) and probe unknown aspects of them. Extensive research has been conducted on these lines in various situations~\cite
	{Morozova13,Bisnovatyi-Kogan2010a,Tsupko12,Cunha20a,Atamurotov21PFDM,Babar21a,Javed22,Jafarzade21a,Atamurotov22,Atamurotov21galaxy}.
	
	To match observational conclusions, it is essential to test the spacetime geometry and study its impact on phenomena occurring in the close vicinity of massive compact objects. In an astrophysical scenario, black holes can only possess mass $M$, rotation $a$, and electric charge $Q$. Among them, the rotation of black holes has been confirmed by a number of observations~\cite
	{Bambi17-BHs,Walton13,Patrick11b-Seyfert,Patrick11a-Seyfert,Tan12,Gallo11}. Reissner-Nordstr\"{o}m (RN) black hole solution is only characterized by mass $M$ and charge $Q$~\cite{Reissner16,Nordstrom18} with interesting properties~\cite{Pugliese11,Pugliese11b}. One of the mechanisms proposed in the literature permits black holes to be charged with a positive net electric charge~\cite{Zajacek19,Bally78}. In addition, the induced field can help black holes to have electric charges under the effect of magnetic field lines~\cite{Wald74}. Different solutions have been proposed in this regard, including a rotating Schwinger dyon black hole solution with electric ($Q_e$) and magnetic ($Q_m$) charges~\cite{Kasuya82,Shaymatov22b}, and regular black hole solutions associated with non-linear electrodynamics (NED)~\cite
	{Bardeen68,Ayon-Beato98,Bronnikov01,Bambi13,Fan2016,Kruglov_2021,Panotopoulos_2020,Rincon2021}. Here, it is worth noting that Einstein's theory of gravity is also applicable for the low energy limit of string field theory that facilitates the dilaton scalar field. Note that this field involves additional terms in the action with the gauge, axion, and dilaton fields~\cite{Green87book,Gibbons88}. For that purpose, the heterotic string theory was also proposed using the scalar dilaton field together with the electromagnetic field~\cite{Garfinkle91}; some representative references in connection with the dilaton fields are~\cite{Gibbons88,Koikawa87,Brill91,Garfinkle91,Rakhmanov94,Harms92}. In the frame of extended theories, black hole solutions have also been considered~\cite{Gubser98,Witten98,Maldacena99,Aharony00,Klemm01} and analyzed with their quantum features  \cite{Rincon_2017,Koch_2014,Bonanno_2021,Koch_2016}). Interesting black hole solutions within the EMS theory involve the dilaton field and cosmological constant~\cite{Gibbons88,Gao04,Yu_2021}. Extensive research in connection with these black hole solutions within the EMS theory has since been devoted to the study of their properties~\cite{Yu_2021,Rayimbaev2020,Zahid2022,Alloqulov24:CPC1,Kurbonov2023}. 
	
	It must be noted that, in previous studies, the dilaton black hole spacetimes can be affected by the dilaton charge, which acts as a new hair as well as a powerful tool for testing optical phenomena around the spacetime. However, in this study, we considered an interesting solution describing a charged black hole within the EMS theory of gravity as the extension of the RN solution involving dilaton field, as described by the line
	element in Eq.~(\ref{Eq:action}), which we can further manifest with details. This solution can directly be affected by the black hole electrical charge, which becomes the main distinguishing characteristics from the dilaton black hole solution. Therefore, it is instructive to thoroughly explore remarkable aspects of such a black hole solution within the EMS theory and the effects of its parameters on the optical properties around the spacetime geometry. This enhances our understanding in relation to its implications in explaining astrophysical observations and distinguishing from other black hole solutions. In this study, we analyzed the optical properties of this black hole solution using shadow and weak gravitational lensing with the magnification of lensed images in the strong field regime. We also analyzed the influence of the EMS black hole parameters on optical phenomena, thus allowing us to gain a deeper understanding of the spacetime geometry. We aim to investigate the shadow of charged black holes within the EMS theory using analytical calculations. A comparison of analytical results with numerical simulations, which can only cover a specific set of parameters, shows that the former illustrate how the impact of these parameters varies and their general characteristics.  
	
	The paper is organized as follows. In Sec.~\ref{Sec:2}, we discuss the metric for a charged black hole within the EMS theory of gravity; this is followed by the study of the black hole shadow and the analytical solutions that provide the radii of the photon sphere and shadow. Sec.~\ref {Sec:3} is devoted to the study of weak gravitational lensing with magnification of lensed images around the black hole. Concluding remarks are discussed in Sec.~\ref{Sec:con}. 
	
	Throughout the paper, we use a system of units in which $G=c=1$ and signature $(-, +, +, +)$ for the metric.

	\section{A charged black hole in Einstein-Maxwell-scalar theory and its impact on black hole shadows\label{Sec:2}}
	
\begin{figure}
		\centering
		\includegraphics[width=8.5cm]{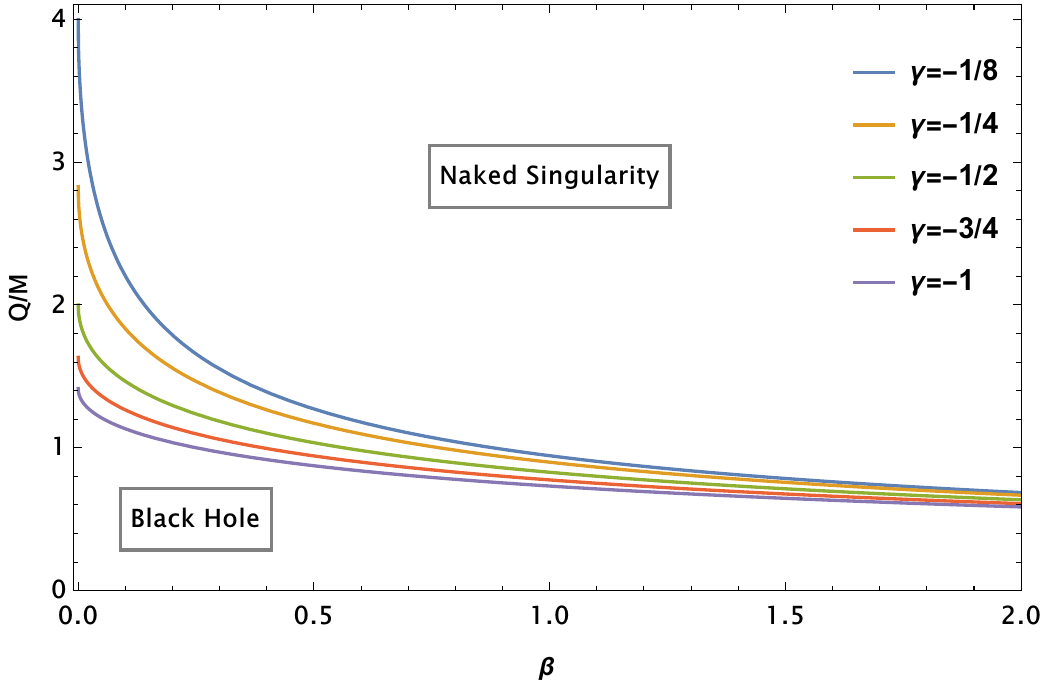}
		\caption{Parameter space plot between the charge parameter $Q/M$ and dimensionless parameter $\beta$ of the black hole within the EMS theory for various combinations of parameter  $\gamma$. }
		\label{Fig:fig1}
	\end{figure}

	The action can be expressed as~\cite{Gibbons88,Yu_2021})
	\begin{eqnarray}\nonumber\label{Eq:action}
		S=\int d^4x\sqrt{-g}\Big[R-2\nabla_\alpha\phi \nabla^\alpha\phi-K(\phi) F_{\alpha\beta}F^{\alpha\beta}
		-V(\phi)
		\Big]\, ,
	\end{eqnarray}
	where new quantities in the action are $\phi$ and $K(\phi)$, which represent the massless scalar field and scalar field function, respectively. It should be noted that $K(\phi)$ is also referred to as the coupling function describing the relation between dilaton fields and the electromagnetic $F_{\alpha\beta}$. The last term $V(\phi)$ in the action denotes the potential pertaining to the cosmological constant $\Lambda$, related to the de-Sitter black hole solution with the dilaton field within the EMS theory; i.e., $V(\phi)=\frac{\Lambda}{3}\left(e^{2\phi}+4+e^{-2\phi}\right)$ \cite{Gao04}. 
	Then, the metric describing a spherically symmetric charged black hole within the EMS theory expressed in Schwarzschild coordinates (i.e., $V(\phi)=0$) is given by~\cite{Yu_2021}
	\begin{equation}\label{metric}
		ds^2=-U(r)dt^2+\frac{dr^2}{U(r)}+f(r)\left(d\theta^2+\sin^2\theta d\varphi^2\right)\, ,
	\end{equation}
	with radial functions $U(r)$ and $f(r)$ for $K(\phi)=\frac{2e^{2\phi}}{\beta e^{4\phi}+\beta-2\gamma}$ expressed as  
	\begin{eqnarray}\label{metfuncts}\nonumber
		&& f(r)=r^2\left(1+\frac{\gamma Q^2}{Mr}\right)\, , \\ && U(r)=1-\frac{2M}{r}+\frac{\beta Q^2}{f(r)}\, .
	\end{eqnarray}
	Note that $M$ and $Q$ are respectively referred to as the mass and electric charge of the black hole, while $\beta$ and $\gamma$ are dimensionless constants within the EMS theory. It should also be noted that $f(r)$ and $U(r)$ can recover the Schwarzschild and Reissner-Nordstr\"{o}m black hole solutions in the case of various combinations of parameter $\beta$ and $\gamma$~\cite{Gibbons88,Garfinkle91}). Moreover, $f(r)$ and $U(r)$ recover the Schwarzschild solution in case the $\beta$ and $\gamma$ parameters are switched off. Similarly, it reduces to the Reissner-Nordstr\"{o}m black hole in the case of $\gamma=0$ and $\beta=1$. However, the above solution turns into the dilation solution when $\beta=0$ and $\gamma=-1$ (see for example \cite{Gibbons88,Garfinkle91}). The black hole horizon $r_{h}$ can be easily determined by setting $U(r)=0$, which is given by 
	\begin{equation}\label{Eq:horizon}
		\frac{r_h}{M}= 1-\frac{\gamma  Q^2}{2 M^2}+\sqrt{1+\frac{Q^2 (\gamma -\beta )}{M^2}+\frac{\gamma ^2 Q^4}{4 M^4}}\, .
	\end{equation}
	Note from the above equation that the black hole horizon no longer exists in the case of a larger parameter $\beta$, thus resulting in the space-time being a naked singularity. We demonstrate it in Fig.~\ref{Fig:fig1}, which plots the parameter space between the charge parameter $Q$ and dimensionless parameter $\beta$ of the black hole for various combinations of the parameter $\gamma$. As can be observed from Fig.~\ref{Fig:fig1}, a black hole sustains its existence in the region which is separated from naked singularity regions by the curves.  We can also approach this issue from a different perspective, i.e., black hole extremes can be determined by imposing the condition $U(r)=U'(r)=0$, thereby obtaining the limiting values of black hole parameters as
	\begin{eqnarray}
		\label{rhmin}
		\frac{(r_h)_{min}}{M}=2-\frac{\beta }{\gamma }+ \frac{\sqrt{\beta ^2-2 \beta  \gamma }}{\gamma }\ ,\\ \label{qextr}
		\frac{Q_{extr}^2}{M^2} =\frac{2 \left(\beta- \sqrt{\beta^2 -2 \beta \gamma }-\gamma \right)}{\gamma ^2}\, .
	\end{eqnarray}  
	
	\begin{figure}
		\includegraphics[scale=0.485]{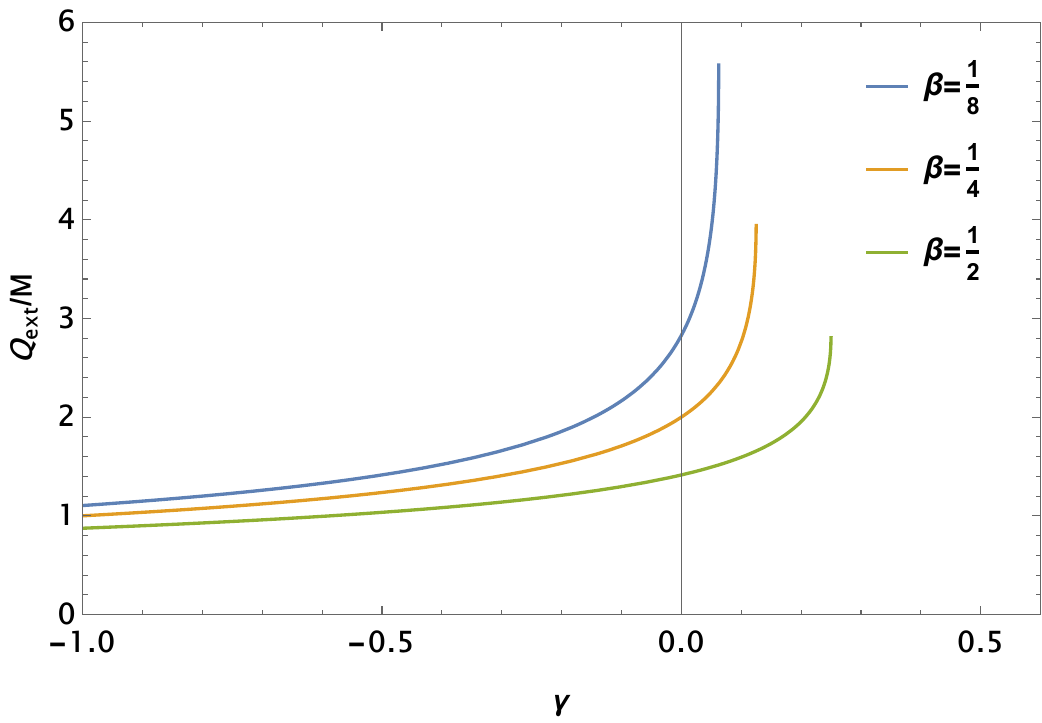}
		\caption{Extreme values of $Q_{extr}/M$ plotted as a function of the parameter $\gamma$ for various combinations of $\beta$.}
		\label{Fig:fig2}
	\end{figure}
	Fig.~\ref{Fig:fig2} represents possible extreme values of $Q$ as a function of $\gamma$ while keeping $\beta$ fixed. As can be observed from Fig.~\ref{Fig:fig2}, extreme values of black hole charge can reach large values as a consequence of an increase in the value of $\gamma$, while the opposite occurs for $\beta$. The extreme conditions Eqs.~(\ref{rhmin}) and (\ref{qextr}) implicitly imply that $\beta_{\rm max}=2 \gamma$ and ${Q_{\rm extr}}/{M}=\sqrt{{2}/{\gamma}}$~\cite{Alloqulov24:CPC1}, addressing the limiting values of black hole parameters.
	
	To study shadow formation, it is necessary to consider the motion of test particles around a static and spherically symmetric black hole solution 
	in the EMS theory metric given by Eq. (\ref{metric}). The Lagrangian corresponding to this metric is
	\begin{equation}
		\mathcal{L}=\frac{1}{2}\left[-U(r)\dot{t}^{2}
		+\frac{1}{U(r)}\dot{r}^{2}+f(r)\left(\dot{\theta}^{2}+\sin^{2}\theta\dot{\varphi}^{2}\right)\right],
		\label{LAG}
	\end{equation}
	To obtain the geodesic equations, we use the following Hamilton-Jacobi equation: 
	\begin{equation}
		\frac{\partial \mathcal{S}}{\partial \sigma }=-\frac{1}{2}g^{\mu \nu }\frac{%
			\partial \mathcal{S}}{\partial x^{\mu }}\frac{\partial \mathcal{S}}{\partial
			x^{\nu }},
	\end{equation}%
	where $\mathcal{S}$ is the Jacobi action. The Jacobi action separable solution reads
	\begin{equation}
		\mathcal{S}=-Et+\ell \varphi +\mathcal{S}_{r}\left( r\right) +\mathcal{S}%
		_{\theta }\left( \theta \right) ,
	\end{equation}%
	where $E$ and $\ell $ are the two Killing vectors of the metric expressed by Eq. (\ref{metric}), given by%
	\begin{equation}
		E=\frac{d\mathcal{L}}{d\overset{\cdot }{t}}=-U\left( r\right) \dot{t}  \label{E1}
	\end{equation}%
	\begin{equation}
		\ell =\frac{d\mathcal{L}}{d\overset{\cdot }{\varphi }}=f(r)\sin ^{2}\theta \overset{\cdot 
		}{\varphi }\text{.}  \label{An1}
	\end{equation}
	Thus, the geodesic equations are
	\begin{equation}
		\frac{dt}{d\sigma }=\frac{E}{U\left( r\right) },\qquad \frac{d\varphi }{d\sigma 
		}=-\frac{\ell }{f(r)\sin ^{2}\theta },
	\end{equation}%
	\begin{equation}
		r^{2}\frac{dr}{d\sigma }=\pm \sqrt{\mathcal{R}\left( r\right) },\qquad r^{2}\frac{d\theta }{d\sigma }=\pm \sqrt{\Theta \left( \theta \right) },  \label{R1}
	\end{equation}%
	where $\mathcal{K}$ is the Carter separation constant and 
	\begin{equation}
		\mathcal{R}\left( r\right) =r^{4}E^{2}-\left( \mathcal{K}+\ell ^{2}\right)
		r^{2}U\left( r\right) ,  \label{R2}
	\end{equation}%
	\begin{equation}
		\Theta \left( \theta \right) =\mathcal{K}-\ell ^{2}\cot \theta .\label{RR2}
	\end{equation}%
	Dimensionless quantities called impact parameters are introduced as 
	\begin{equation}
		\eta =\frac{\mathcal{K}}{E^{2}},\qquad \text{ }\zeta =\frac{\ell }{E}.
		\label{impact1}
	\end{equation}%
	It depends on the values of critical parameters whether the photon is
	captured, scattered to infinity, or bound to orbits. Our interest is in spherical light geodesics constrained on a sphere of constant coordinate
	radius $r$ with $\overset{\cdot }{r}=0$ and $\overset{\cdot \cdot }{r}=0$, also known as spherical photon orbits. Without any loss of
	generality, we set the equatorial plane $\theta=\pi/2$. Circular orbits correspond to the
	maximum effective potential, and unstable photons must satisfy the
	following conditions:%
	\begin{equation}
		V_{eff}\left( r\right) \left\vert _{r=r_{ps}}\right. =0,\hspace{1cm}V_{eff}^{\prime
		}\left( r\right) \left\vert _{r=r_{ps}}\right. =0,
	\end{equation}%
	or,%
	\begin{equation}
		\mathcal{R}\left( r\right) \left\vert _{r=r_{ps}}\right. =0,\hspace{1cm}\mathcal{R}%
		_{eff}^{\prime }\left( r\right) \left\vert _{r=r_{ps}}\right. =0.  \label{R3}
	\end{equation}%
	where $r_{ps}$ is the photon sphere and determines the location of the apparent
	image of the photon rings; the Carter separation constant disappears on recasting Eqs. (\ref{R2}) and (\ref{RR2}). If we consider the metric expressed by Eq. (\ref{metric}) then we can express the radius of the photon sphere as the solution of the equation 
	\begin{equation}
		f^{\prime }\left( r_{ps}\right) U\left( r_{ps}\right) -f\left( r_{ps}\right)
		U^{\prime }\left( r_{ps}\right) =0,
	\end{equation}%
	or explicitly%
	\begin{equation*}
		6M^{4}r^{2}-2M^{3}\left( r^{3}+Q^{2}r\left( 2\beta -5\gamma \right) \right)
	\end{equation*}
	\begin{equation}
		+M^{2}Q^{2}\left( -3r^{2}-2Q^{2}\left( \beta -2\gamma \right) \gamma
		-Q^{4}r\gamma ^{2}\right) =0. \label{rps1}
	\end{equation}%
	The above equation is a cubic equation, that is, it has three analytical roots. Using the Mathematica 11 software, the only real root is given by
	\begin{equation*}
		r_{ps}=\frac{2M^{2}-Q^{2}\gamma }{2M}+\frac{A}{2^{2/3}3M^{2}\left( B+\sqrt[3]%
			{4A^{3}+B^{2}}\right) ^{1/3}}-
	\end{equation*}\begin{equation}
		\frac{\left( B+\sqrt[3]{4A^{3}+B^{2}}\right)
			^{1/3}}{2^{1/3}6M^{2}},\label{rps2}
	\end{equation}%
	where%
	\begin{equation*}
		A=3M^{2}\left( -12M^{4}+6M^{2}Q^{2}\left( \beta -\gamma \right) -Q^{4}\gamma
		^{2}\right)\end{equation*} and \begin{equation*}
		B=108M^{5}\left( -4M^{4}+4M^{2}Q^{2}\left( \beta -\gamma
		\right) -Q^{4}\gamma ^{2}\right) .
	\end{equation*}%
	\begin{figure}
		\centering
		\includegraphics[width=8.5cm]{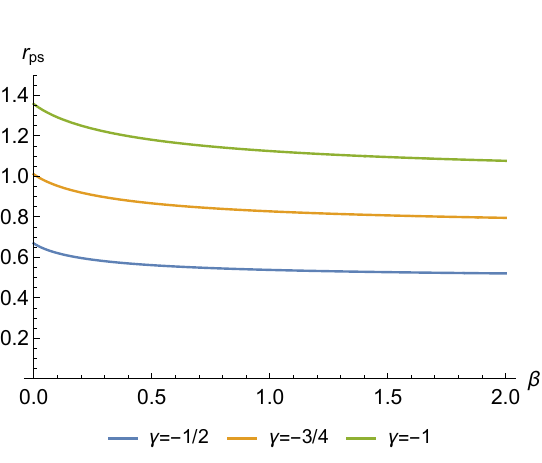}
		\caption{Plot of Eq. (\ref{rps2}) illustrating the dependence of $r_{ps}$ on the black hole parameters $\beta$ and $\gamma$; here, $Q/M=0.66$.}
		\label{figa1}
	\end{figure}\begin{figure}
		\centering
		\includegraphics[width=8.5cm]{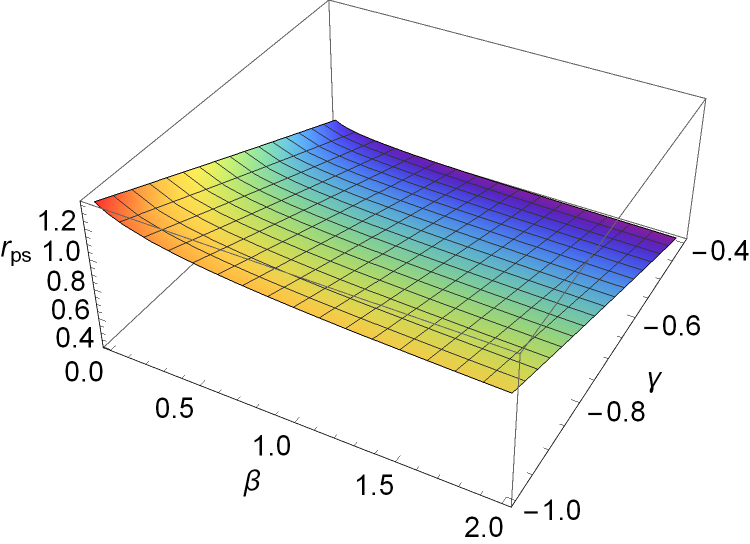}
		\caption{Three-dimensional plot of Eq. (\ref{rps2}) illustrating the dependence of $r_{ps}$ on the black hole parameters $\beta$ and $\gamma$; here, $Q/M=0.66$.}
		\label{figa2}
	\end{figure}
	To illustrate this, Fig. \ref{figa1} represents the radius of photon sphere with respect to the parameters $\beta$ and $\gamma$. As can be seen from Fig. \ref{figa1}, the photon sphere increases as the magnitude of the $\gamma$ parameter
	increases. However, as the parameter $\beta$ increases, the photon sphere
	decreases first and then remains constant regardless of how much it
	increases. Figure \ref{figa2} depicts a three-dimensional plot of the radius of the
	photon sphere with respect to the parameters $\beta $ and $\gamma$, revealing
	their effect on $r_{ps}$.
	\begin{figure}
		\centering
		\includegraphics[width=8.5cm]{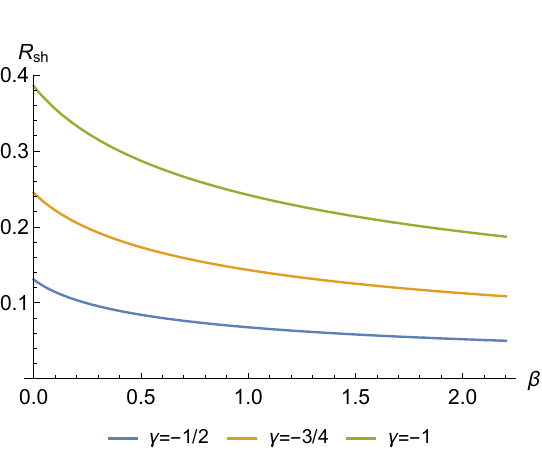}
		\caption{Variation of the shadow observable $R_{sh}$ according to Eq. (\ref{rshadow}) for the charged black hole within the EMS theory; here, $Q/M=0.66$.}
		\label{figa3}
	\end{figure}\begin{figure}
		\centering
		\includegraphics[width=8.5cm]{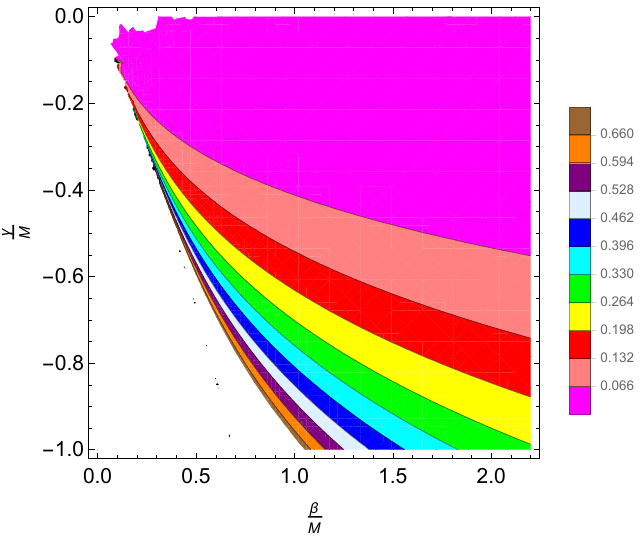}
		\caption{Variation of the contour plot for the charged black hole within the EMS theory; here, $Q/M$=0.66.}
		\label{figa4}
	\end{figure}

\begin{figure*}
		\centering
		{{\includegraphics[width=7.5cm]{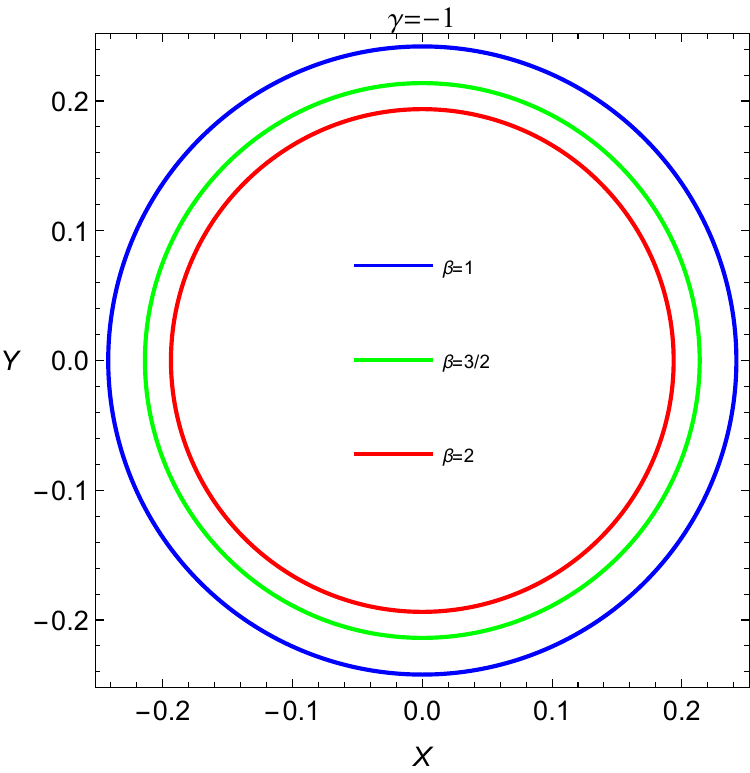} }}\qquad
		{{\includegraphics[width=7.5cm]{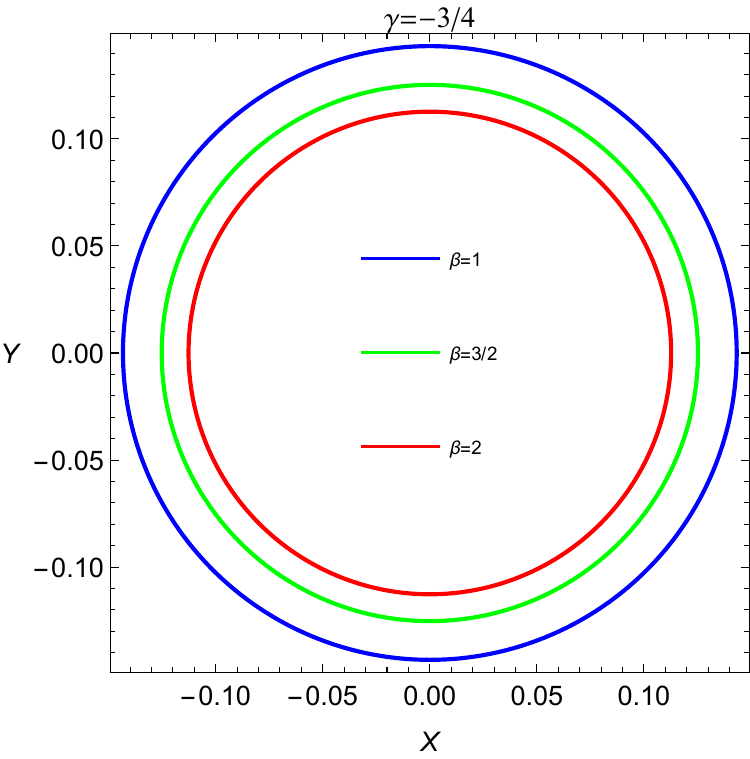}}}
		\caption{Profile of shadows cast by the charged black hole within the EMS theory for different values of $\beta$ and $\gamma$; here, $Q/M=0.66$.}
		\label{figa5}
	\end{figure*}

	Accordingly, the radius of the shadow $R_{sh}$ is defined by the lensed image of the photon sphere as
	\begin{eqnarray}\label{rshadow}
		&& R_{sh}=\left. \sqrt{\frac{f\left( r_{ps}\right) }{U\left( r_{ps}\right) }} 
		\right\vert _{r=r_{ps}} \\ \nonumber 
		&&
		=\sqrt{-\frac{r_{ps}^2(Mr_{ps}+Q^2\gamma)^2 }{M(2M^2r_{ps}-M(r_{ps}^2+Q^2(\beta-2\gamma))-Q^2r_{ps}\gamma)}}.
	\end{eqnarray}
	This coincides with the value of the impact parameter itself. Figures \ref{figa3} and \ref{figa4}
	depict the variation of the shadow observable $R_{sh}$ and the contours plot, respectively,  
	for the charged black hole solution within the EMS theory in
	the ($\beta $, $\gamma $ ) space. They show how the shadow size
	varies. Note that an increase in the magnitude of the $\gamma$ parameter leads to an increase in the size of the black hole
	shadow. In contrast, the $\beta$ parameter has the opposite effect: it decreases the shadow of the black hole.  
	\begin{table} 
		\begin{center}
			\resizebox{.5\textwidth}{!}
			{
				\begin{tabular}{|c|c|c|c|c|c|c|}  \hline
					&  \multicolumn{2}{|c|}  {$\gamma =-1/2$} &  \multicolumn{2}{|c|}  {$\gamma =-3/4$}
					& \multicolumn{2}{|c|} {$\gamma =-1$}   \\ \hline  
					& $r_{ps}/M$ & $R_{sh}/M$ & $r_{ps}/M$ & $R_{sh}/M$ & $r_{ps}/M$ & $R_{sh}/M$
					\\ \hline
					$\beta =1$ & 0.5377  & 0.0678  & 0.8268 & 0.1433  & 1.124 & 
					0.24206 \\ 
					$\beta =3/2$ & 0.52669 & 0.05842 & 0.80688 & 0.12522  & 1.0949 & 
					0.21377 \\ 
					$\beta =2$ & 0.52024  & 0.05210 & 0.79469 & 0.11269 & 1.0762 & 0.19371%
					\\ \hline
				\end{tabular}
			}
			\caption{Numerical results for the values of $r_{ps}$ and  $R_{sh}$ for a black hole within the EMS theory; here, $Q/M=0.66$.}
			\label{taba1}
		\end{center}
	\end{table}\\
	Celestial coordinates are used to describe the shadow of the black hole seen on an observer's frame \cite{vazquez}. Thus, we define the celestial coordinates  $X$ and $Y$ by \begin{equation}
		X=\lim_{r_{0}\rightarrow \infty }\left( -r_{0}\sin \theta _{0}\left. \frac{
			d\varphi }{dr}\right\vert _{r_{0},\theta _{0}}\right) ,  \label{x11}
	\end{equation}
	\begin{equation}
		Y=\lim_{r_{0}\rightarrow \infty }\left( r_{0}\left. \frac{d\theta }{dr}
		\right\vert _{r_{0},\theta _{0}}\right) ,  \label{y11}
	\end{equation}
	where $(r_{0},\theta_{0})$ are the position coordinates of the observer. Assuming that the observer is on the equatorial hyperplane, Eqs. 
	(\ref{x11}) and (\ref{y11}) follow \begin{equation}
		X^{2}+Y^{2}=R_{sh}^{2}.  \label{xy1}
	\end{equation} 
	Table \ref{taba1} lists the numerical values of $r_{ps}$ and $R_{sh}$ for a specific set of parameters. The profile of the shadows cast by the charged black hole within the EMS theory is shown in Fig.~\ref{figa5}  under the influence of the parameters $\beta$  and $\gamma$. Figure  \ref{figa5} clearly shows that the shadow radii decrease in black holes as $\beta$ increases and that the decrements of the shadow radii also increase with different intervals.

	\section{Weak gravitational lensing and magnification of lensed image}\label{Sec:3}
	
	\begin{figure*}
		\centering
		\includegraphics[scale=0.4]{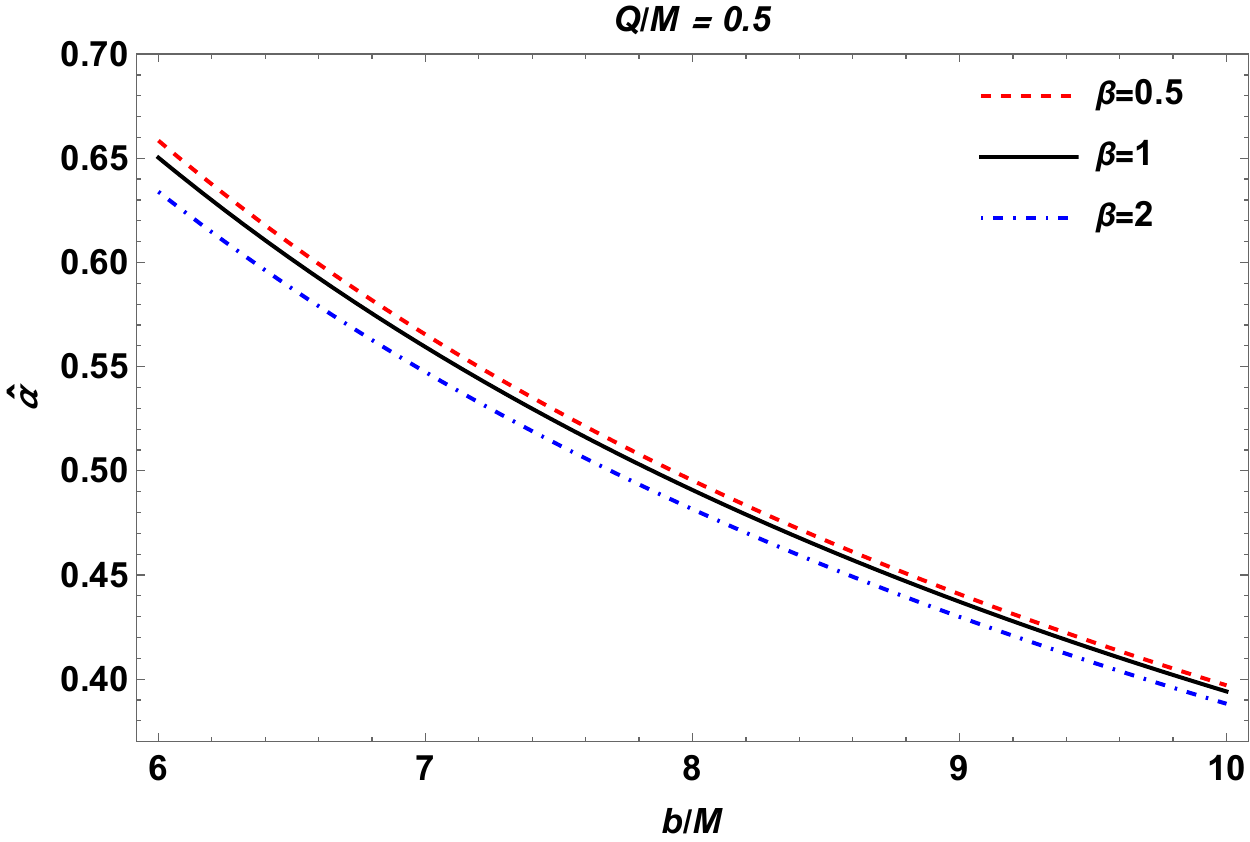}
		\includegraphics[scale=0.4]{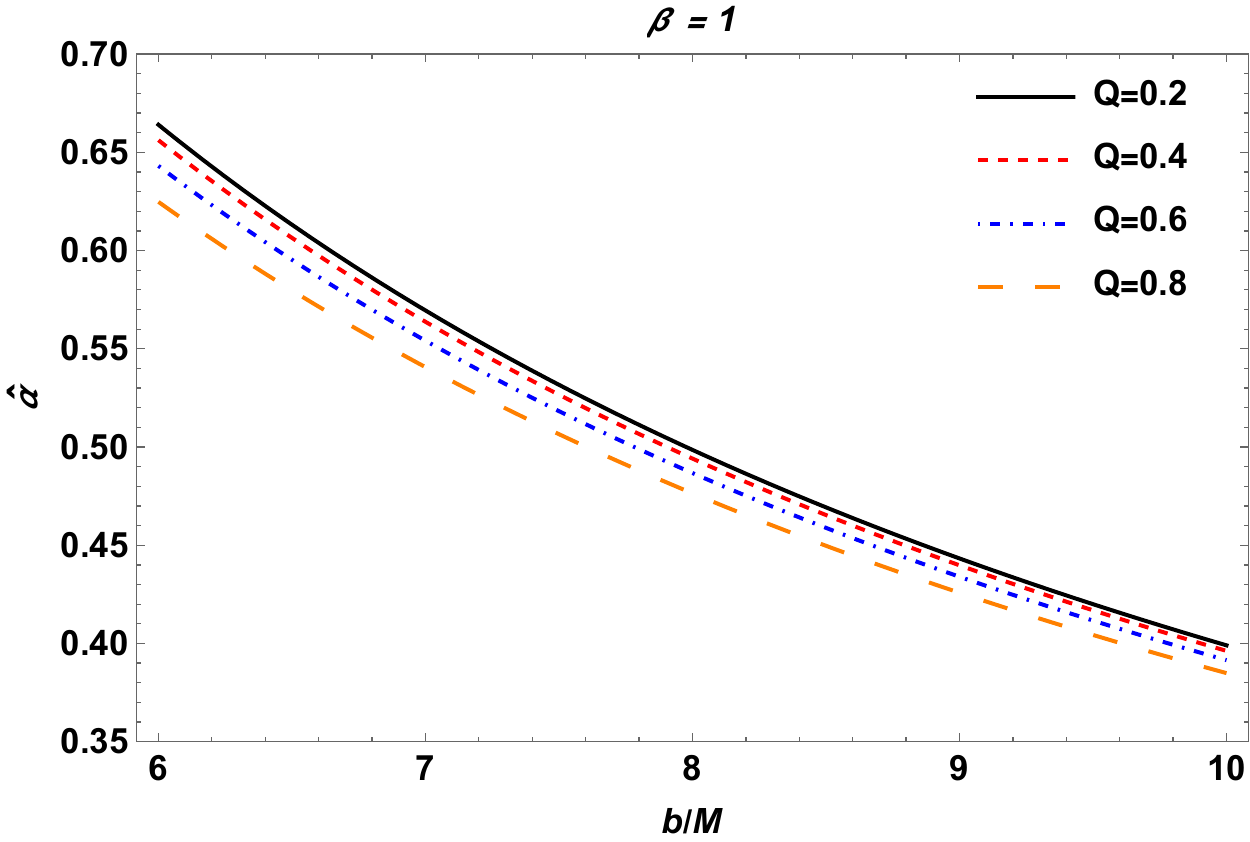}
		\caption{Deflection angle $\hat{\alpha}$ plotted as a function of the impact parameter $b$ for different combinations of parameter $\beta$ (left panel) and black hole charge (right panel) for a fixed value of $\gamma$.}
		\label{fig:alpha}
	\end{figure*}
	
	In this section, we examine the weak gravitational lensing around the black hole within the EMS theory. Interestingly, the deviation of a ray of light from its original path occurs when it passes through the close vicinity of massive objects. 
	For a weak-field approximation, the following relation can be used as a metric tensor:
	\begin{equation}
		g_{\alpha \beta}=\eta_{\alpha \beta}+h_{\alpha \beta}\, ,
	\end{equation}
	where $\eta_{\alpha \beta}$ and $h_{\alpha \beta}$ refer to expressions for the Minkowski spacetime and
	perturbation gravity field describing the EMS theory, respectively. For the weak gravitational field to be satisfied, the following expressions for $\eta_{\alpha \beta}$ and $h_{\alpha \beta}$ must be satisfied: 
	\begin{eqnarray}
		&&   \eta_{\alpha \beta}=diag(-1,1,1,1)\ , \nonumber\\
		&&   h_{\alpha \beta} \ll 1, \hspace{0.5cm} h_{\alpha \beta} \rightarrow 0 \hspace{0.5cm} \mbox{under} \hspace{0.2cm}  x^{\alpha}\rightarrow \infty \ ,\nonumber\\
		&&     g^{\alpha \beta}=\eta^{\alpha \beta}-h^{\alpha \beta}, \hspace{0,5cm} h^{\alpha \beta}=h_{\alpha \beta}\, ,
	\end{eqnarray}
	where $x^{\alpha}$ refers to the spacetime coordinate. 
	
	Using the fundamental equation, we can obtain the expression for the deflection angle around a compact object within EMS theory as follows: 
	\begin{equation}
		\hat{\alpha }_{\text{b}}=\frac{1}{2}\int_{-\infty}^{\infty}\frac{b}{r}\left(\frac{dh_{33}}{dr}+\frac{dh_{00}}{dr}\right)dz\ , 
	\end{equation}
	We can express the line element in Eq.~\ref{metric} as follows:
	\begin{eqnarray}
		ds^2 \approx ds_0^2+ \Big(\frac{2M}{r}-\frac{\beta Q^2}{f(r)}\Big)dt^2+ \Big(\frac{2M}{r}-\frac{\beta Q^2}{f(r)}\Big)dr^2
	\end{eqnarray}
	where $ds^2_0=-dt^2+dr^2+r^2(d\theta^2+\sin^2\theta d\phi^2)$.
	We can easily find components $h_{\alpha \beta}$ of the metric element in Cartesian coordinates as follows:
	\begin{eqnarray}
		h_{00}&=&\frac{2M}{r}-\frac{\beta Q^2}{f(r)}\\
		h_{ik}&=&\Big(\frac{2M}{r}-\frac{\beta Q^2}{f(r)}\Big)n_i n_k \\
		h_{33}&=&\Big(\frac{2M}{r}-\frac{\beta Q^2}{f(r)}\Big) \cos^2\chi \label{h}\ ,
	\end{eqnarray}
	where $\cos^2\chi=z^2/(b^2+z^2)$ and $r^2=b^2+z^2$.\\
	Now, we can define the derivatives of $h_{00}$ and $h_{33}$ using radial coordinates. Subsequently, we can calculate the deflection angle $\hat{\alpha}_b$.

	Accordingly, we determine an explicit form of the deflection angle analytically. To this end, we restrict the location of the observer to the equatorial plane, i.e., $\theta=\pi/2$. To analyze the deflection angle, we use the Hamilton formalism to evaluate the geodesic equations. The standard Hamiltonian is expressed as    \begin{equation}\label{Hamiltonian}
		H(x,p)=\frac{1}{2}g^{\alpha \beta}(x)p_{\alpha}p_{\beta}\, ,
	\end{equation}
	with 
	\begin{eqnarray}
		\dot{p_{\alpha}}=-\frac{\partial H}{\partial x^{\alpha}}  \qquad  \mbox{and} \qquad \dot{x^{\alpha}}=\frac{\partial H}{\partial p_{\alpha}}\, .
	\end{eqnarray}
	From the Hamilton-Jacobi equation, we can also derive
	\begin{align}
		\dot{\varphi}=\frac{\partial H}{\partial p_{\varphi}}=g^{\varphi \varphi}p_{\varphi} \qquad  \mbox{and} \qquad \dot{r}=\frac{\partial H}{\partial p_r}=g^{rr}p_{r}\, .
	\end{align}
	From the above equations, a simplified form is given by
	\begin{equation}
		\left(\frac{\dot{r}}{\dot{\varphi}}\right)^2=\left(\frac{g^{rr}p_r}{g^{\varphi \varphi}p_{\varphi}}\right)^2\, .
	\end{equation}
	It should be noted that the Hamiltonian can be considered as $H=0$ for the null particle. Thus, Eq.~(\ref{Hamiltonian}) can be rewritten on the basis of $p_t=-E$ and $p_{\varphi}=l$, i.e., 
	\begin{equation}
		g^{rr}p^2_r=-(g^{tt}E^2+g^{\varphi \varphi}l^2)\, ,
	\end{equation}
	leading to
	\begin{eqnarray}\label{Eq r/phi}
		&& \left(\frac{\dot{r}}{\dot{\varphi}}\right)^2=-\frac{g^{rr}}{(g^{\varphi \varphi}l)^2}(g^{tt}E^2+g^{\varphi \varphi}l^2)\, .
	\end{eqnarray}
	Taking $b={E/}{l}$ into consideration, which is referred to as the impact parameter, we can rewrite Eq.~(\ref{Eq r/phi}) as 
	\begin{equation}\label{Eq:r/phi2}
		\left(\frac{\dot{r}}{\dot{\varphi}}\right)^2=-\frac{g^{rr}}{(g^{\varphi\varphi})^2}(g^{tt}b^2+g^{\varphi\varphi})\, .
	\end{equation}
	As a matter of fact, the deviation of the light ray leads to the deflection angle by which the ray is bent from its original path when passing through a massive object. Hence, the deflection angle can be evaluated when the light
	is bent from its original path at the closest distance from the massive object (i.e., $r = r_0$). Furthermore, to determine the impact parameter at $r=r_0$, one can set the following condition:
	\begin{equation}\label{Eq:r/phi3}
		\left(\frac{\dot{r}}{\dot{\varphi}}\right)\Big|_{r=r_0}=0\, , 
	\end{equation}
	with 
	\begin{eqnarray}\label{Eq:G}
		g^{tt}|_{r=r_0}=G^{tt},\quad g^{\varphi\varphi}|_{r=r_0}=G^{\varphi\varphi}, \quad g^{rr}|_{r=r_0}=G^{rr}\, .
	\end{eqnarray}
	Considering Eqs.~(\ref{Eq:r/phi3})and (\ref{Eq:G}) together, the impact parameter can be expressed as follows: 
	\begin{equation}\label{Eq:lambda}
		b^2=-\frac{G^{\varphi\varphi}}{G^{tt}}\, .
	\end{equation}
	According to Eqs.~(\ref{Eq:r/phi2}) and (\ref{Eq:lambda}), 
	the integral form of the deflection angle by which the light is deviated from its original path can be defined as  
	\begin{equation}\label{defangle}
		\int_0^{\bar{\alpha}} d \varphi=\pm 2 \int_{-\infty}^{\infty} \Big[\frac{-g^{rr}}{(g^{\varphi \varphi})^2}\left(g^{tt}b^2+g^{\varphi \varphi}\right)\Big]^{-1/2}dr\, .
	\end{equation}
	The important point to be noted here is that one can take into account $\pi$ when evaluating the deflection angle of the light ray deviating from its original trajectory if and only if the coordinate's center refers to the compact object.   
	Accordingly, the real deflection angle by which the light is bent from its original path can be defined as 
	$\hat{\alpha_b}=\bar{\alpha}-\pi$. However, the analytical integration Eq.~(\ref{defangle}) for the deflection angle is complicated. Therefore, we resort to the numerical evaluation of the deflection angle $\hat{\alpha_b}$. To gain a deeper understanding on the deflection angle of the light ray, we analyzed its behaviour. Its dependence on the impact parameter for various values of black hole parameters is shown in Fig.~\ref{fig:alpha}. Note from Fig.~\ref{fig:alpha} that the deflection angle of the light ray decreases with the increase in the impact parameter $b/M$, whereas the curves shift downwards to smaller values with the increase in the black hole charge and parameter $\beta$.
	
\begin{figure}
		\centering
		\includegraphics[scale=0.4]{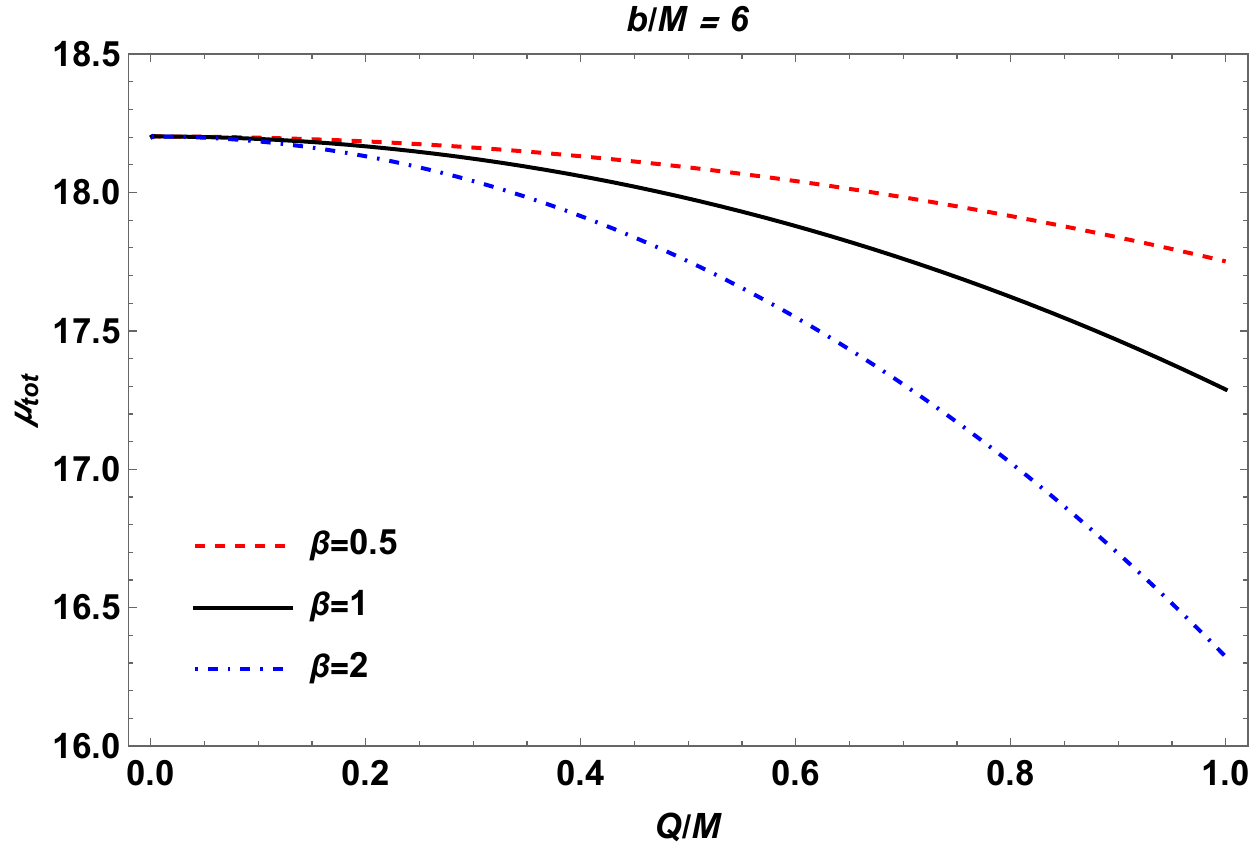}
		\caption{Total magnification $\mu_{tot}$ plotted as a function of the black hole charge for different values of $\beta$ and a fixed value of $\gamma$. Note that we set the impact parameter as $b = 6M$.}
		\label{fig:magnification}
	\end{figure}

	Let us now examine the brightness of the image using the light's deflection angle around the black hole within the EMS theory. To this end, let us consider the following expression, given in terms of the light angles, such as $\hat{\alpha_b}$, $\theta$ and $\beta$ \cite{Morozova13,Bozza2008lens,Babar21a}): 
	\begin{eqnarray}\label{lenseq}
		\theta D_\mathrm{s}=\beta D_\mathrm{s}+\hat{\alpha_b}D_\mathrm{ds}\, . 
	\end{eqnarray}
	Note that in Eq.~(\ref{lenseq}) we represent the distances between the source and observer, $D_\mathrm{s}$, the lens and the observer, $D_\mathrm{d}$, and the source and the lens, $D_\mathrm{ds}$, whereas $\theta$ and $\beta$ denote the angular position of the image and source, respectively. Based on Eq.~(\ref{lenseq}), the angular position $\beta$ of the source is expressed as   
	\begin{eqnarray}\label{newlenseq}
		\beta=\theta -\frac{D_\mathrm{ds}}{D_\mathrm{s}}\frac{\xi(\theta)}{D_\mathrm{d}}\frac{1}{\theta}\, .
	\end{eqnarray}
	Note that we have used $\xi(\theta)=|\hat{\alpha}_b|\,b$, with $b=D_\mathrm{d}\theta$ ~\cite{Bozza2008lens}. Accordingly, one can determine the shape of the image as Einstein ring using the radius $R_s=D_\mathrm{d}\,\theta_E$, provided that its shape behaves like a ring. In Eq.~(\ref{newlenseq}), the angular part $\theta_E$, which appears owing to spacetime geometry between the source images, can be expressed as~\cite{Morozova13}
	\begin{eqnarray}
		\theta_E=\sqrt{2R_s\frac{D_{ds}}{D_dD_s}}\, .
	\end{eqnarray}
	Next, we examine the magnification of brightness defined as~\cite{Babar21a,Atamurotov2022,Far:2021a,Alloqulov_2023,10.1088/1674-1137/ad1677}
	\begin{eqnarray}\label{magni}
		\mu_{\Sigma}=\frac{I_\mathrm{tot}}{I_*}=\underset{k}\sum\bigg|\bigg(\frac{\theta_k}{\beta}\bigg)\bigg(\frac{d\theta_k}{d\beta}\bigg)\bigg|, \quad k=1,2, \cdot \cdot \cdot ,  j\, ,\nonumber\\
	\end{eqnarray}
	where $I_\mathrm{tot}$ denotes the total brightness whereas $I_*$ denotes the unlensed brightness of the source. The total magnification can thus be expressed as 
	\begin{eqnarray}\label{magtot}
		\mu_\mathrm{tot}=\frac{x^2+2}{x\sqrt{x^2+4}}\, .
	\end{eqnarray}
	Here, $x={\beta}/{\theta_E}$ refers to a dimensionless quantity. Let us now explore the magnification of the source numerically. In particular, we analyzed the dependence of the total magnification on the black hole electric charge for different values of the parameter $\beta$, as shown in Fig.~\ref{fig:magnification}. Note from Fig.~\ref{fig:magnification} that the total magnification decreases with the increase in the black hole charge, whereas its curves shift downwards to smaller values as a consequence of an increase in the value of the parameter $\beta$.

	\section{Conclusions}\label{Sec:con}
	
	Optical studies of black holes play a crucial role for providing information in connection with distance sources, compact objects, and their fascinating nature. Hence, we considered the charged black hole solution within the EMS theory of gravity to understand its spacetime geometry. The strong field regime can have a significant impact not only on massive particle geodesics but also on the null geodesics, which can give rise to a change in observable quantities, including the radius of the shadow $R_{sh}$. Therefore, it is important to study the impact of the spacetime geometry on observable quantities in the close vicinity of black holes, thus resulting in certainty with respect to observational conclusions. 
	
	In this study, we analyzed the optical properties, e.g., the photon motion and weak gravitational lensing around the black holes within the EMS theory for various situations. We calculated the shadow cast by a black hole. We obtained analytical solutions for both the radius of the photon sphere and that of the shadow. Our results show that the black hole parameters $\gamma$ and $\beta$ both influence the shadow of black holes. The radius of the photon sphere and that of the shadow both increase as the magnitude of the parameter $\gamma$  increases. However, as the parameter $\beta$ increases, both radii decrease first and then remain constant, no matter how much $\beta$ increases (Figs. \ref{figa2} and \ref{figa3}). According to the obtained results, the size of the shadow (Fig. \ref{figa5}) of the charged black hole within the EMS theory is highly dependent on the parameter $\beta$, and for large values of $\beta$, the shadow size is reduced significantly.

	We also investigated the weak gravitational lensing for the black hole within the EMS theory. In this regard, we calculated the deflection angle of the light in the weak field regime. Furthermore, we represented the dependence of the deflection angle on the impact parameter for different values of the black hole parameters; see Fig.~\ref{fig:alpha}. We inferred from the result that the value of the deflection angle decreases with the increase in the impact parameter, black hole charge, and parameter $\beta$.
	Finally, we studied the total magnification of the images. The dependence of the total magnification on the black hole charge was demonstrated in Fig.~\ref{fig:magnification}. The total magnification clearly decreases with the increase of the black hole charge and parameter $\beta$. 
	
	These theoretical analyses provide information in connection with black holes within the EMS theory of gravity to explain astrophysical observations.

	\section{Acknowledgments}
	The research is supported by the National Natural Science Foundation of China under Grant No. 11675143 and the National Key Research and Development Program of China under Grant No. 2020YFC2201503. M.A and B.A wish to acknowledge the support from Research Grant F-FA-2021-432 of the Ministry of Higher Education, Science and Innovations of the Republic of Uzbekistan.
	
	\bibliographystyle{apsrev4-2}
	\bibliography{ref}
	
\end{document}